# How come the Correlations?

Nicolas Gisin
Group of Applied Physics
Geneva University
1211 Geneva 4

In relativity there is space-time out there. In quantum mechanics there is entanglement. Entanglement manifests itself by producing correlations between classical events (e.g. the firing of some detectors) at any two space-time locations. If the locations are time-like separated, i.e. one is in the future of the other, then there is no specific difficulty to understand the correlations. But if the two locations are space-like separated, the problem is different. How can the two space-time locations out there know about what happens in each other without any sort of communication? If space-time really exists, the locations must do something like communicating. Or it was all set up at the Beginning. But the correlations depend also on the free choice of the experimentalists, one in each space-time location. This allowed John Bell to derive his inequality and the experimentalists to violate it, thus refuting the assumption that it was all set up at the beginning: the Correlations can't be explained by common causes.

Consequently, either space-time is an illusion, or free will is an illusion, or there is some communication. We call this hypothetical "message" quantum information. It is important to note that quantum information is not under our control, hence this sort of communication can not be used to send classical messages. There is thus no straightforward conflict with relativity.

In the following we shall mainly concentrate on the third alternative (i.e. assume that space-time is real – and free will not an illusion – but that there is some communication), motivated by the possibility to experimentally test some of its predictions.

If space is real, then quantum information can be said to have a velocity, denoted $v_{QI}$ [1].

- If $v_{QI} \leq c$ (c=speed of light), then one would not observe correlations due to entanglement between space-like separated locations, because the information would not get there on time. This is refuted by experiments.

- But $v_{QI} > c$ makes sense only if there is a preferred reference frame. This frame could be either a universal preferred frame, like, for instance, the one determined by the cosmic microwave background radiation, or it could be determined case by case by the very condition of the experiment. In the first case the Correlations are universal, only limited by the speed $v_{QI}$: locations inside a radius get the quantum information on time, outside locations don't; hence the latter show no Correlations (at least none due to entanglement). In the second case, the presence or absence of Correlations depends on the precise experimental condition. Both cases can be experimentally investigated. We shall develop this further below.

- If $v_{QI}$ is infinite, then quantum information is simultaneously everywhere, in all reference frames. In this case space-time is not really out there, but seems to be part of the quantum state of the Universe. One would have to explain why there is apparent space-time? And apparent locality? And why do we sense free will? Indeed, we would also be part of the Universal quantum state, obeying some sort of Schrödinger equation.

In order to test the assumption that $v_{QI}$ is finite but larger than the speed of light, we performed several experiments in Geneva. The results can be found in a series of papers [2-6]. Here we simply and very briefly present the general idea and results.

In a first series of experiments we exploited installed telecom optical fibers (provided by Swisscom) to investigate quantum correlations between two villages separated by more than 10 km in straight line [7]. We made special efforts to guarantee that, in the natural reference frame determined by the Geneva lake and the Swiss Alps, the two events (one in each village) happened in coincidence, with a time precision of 5 ps (corresponding to 1 mm of optical fiber, i.e. a relative precision of $10^{-7}$). The quantum correlations could still be observed, with a visibility large enough for a potential violation of Bell's inequality. Consequently, the speed of quantum information (if it exists!) must be larger than about $10^7 \cdot c$ in the "natural" reference frame. There is clearly no reason to believe that the Geneva reference frame is a universal one, but under the assumption that the relevant frame is defined by the "very experimental conditions", this seems a very natural one. We also analyzed the above experiment as seen from the "center of mass of the universe", that is from the reference frame in which the microwave background radiation is most isotropic [8]. Because of the earth rotation and of the time required to register a full 2-photon interference fringe, the precision we achieved is not as high. Still, this analysis sets a limit to $v_{QI}$ in this universal frame of about $10^4 \cdot c$ ! These figures are very large indeed and most physicists would be tempted to jump to the conclusion that they demonstrate that there is nothing like a speed of quantum information. But, considering the other two alternatives, it might still be worth pursuing this line of research. There is one (actually probably more than one) hidden assumption in the above result: the experimentalist had to decide what to align! Should it be the detector's surfaces? The beam splitters? The region of the avalanche photodiode where the avalanche happens? The electronics? Or something else? In the above experiment we decided to align the surfaces of the detectors, since it is there (we believe) that the irreversible choice happens. But we would be happy to read about alternative (feasible!) suggestions.

In a second series of experiment, following the intuition of Antoine Suarez and Valerio Scarani [9], we explored the consequence of the assumption that the relevant reference frames are determined locally at each side by the "very condition of the experiment". More precisely, the intuition is that if the observers at both ends of the experiment are in relative motion such that each one in its own reference frame is first to do the measurement, then the correlations should disappear. This is not a very precise idea, but actually a quite intuitive one (with all the danger of intuition in a quantum+relativity context [10]). Now, it is a beautiful idea precisely because it allows one to test experimentally a rather natural intuition. Moreover, at the time the proposal was made, no such test had ever been performed. As in the first series, the experimentalist had to decide what to set in motion. Indeed, it is unpractical to imagine an entire lab moving! In a first experiment, the idea has been to set the detectors in motion. But even this turned out to be very challenging. Hence the detectors where replaced by absorbers (the choice being photon absorbed vs. photon not absorbed) and the results read by a photon counter mounted on the second port of the interferometers. The "real detector" was mounted such that the absorber was in its absolute past (i.e. the detector merely reveals the choice made at the absorber). In a second experiment, we decided to set the beam splitters in motion (much in the spirit of the de Broglie-Bohm pilot wave model). Both experiments revealed the presence of quantum correlations, even under these weird conditions.

Most physicists are happy with our experimental results: they conclude that quantum theory is once again well supported by experimental data. (Some will even claim that the experiments where not necessary since they know that quantum theory is correct!). However, the issue is not a matter of happiness or of simple belief in a theory! If the speed of quantum information is indeed infinite, or non-existing, then we are left with the two remaining alternatives: either space-time or free will is an illusion. I am tempted to vote for the first one! But – again – it is not a matter of personal preference. The real problem for physics is the following: how could one test it?

The question "How come the Correlation?" is not part of mainstream physics. But I strongly believe that it is one of the most – possibly the most – important for physics today.

*Acknowledgements*: This work would not have been possible without the financial support of the "Fondation Odier de psycho-physique" and the Swiss National Science Foundation.